%% file: main.tex
\documentclass[10pt]{article}
\usepackage{mathtools}
\usepackage{graphicx}
\usepackage{xcolor}
\usepackage{enumitem}
\usepackage{amsmath}
\usepackage{amssymb}
\usepackage{tikz-cd}
\usepackage{xspace}
\usepackage{mathtools}
\usepackage{algorithm}
\usepackage{algpseudocode}
\usepackage{subcaption}
\usepackage{longtable}
\usepackage{array}
\usepackage[colorlinks=true, allcolors=blue]{hyperref}
\usepackage{authblk}
\usepackage[top=1.2in, bottom=1.2in, left=1.71in, right=1.71in]{geometry}
\input{dfn}
\usepackage{adjustbox}
\usepackage{booktabs}
\usepackage{titlesec}
\usepackage{multirow}

\titleformat{\section}
{\normalfont\fontsize{14}{18}\selectfont\bfseries}{\thesection}{1em}{}
\titleformat{\subsection}
{\normalfont\fontsize{13}{17}\selectfont\bfseries}{\thesubsection}{1em}{}
\titleformat{\subsubsection}
{\normalfont\fontsize{12}{16}\selectfont\bfseries}{\thesubsubsection}{1em}{}

\newtheorem{definition}{Definition}

\title{Encrypted Large Model Inference: The Equivariant Encryption Paradigm}

\author{\vspace{0.25cm}\textbf{\small{James Buban, Hongyang Zhang, Claudio Angione, Harry Yang, Ahmad Farhan, Seyfal Sultanov, Michael Du, Xuran Ma, Zihao Wang, Yue Zhao, Arria Owlia, Fielding Johnston, Patrick Colangelo}}\vspace{0.25cm}}

\affil{\large{Nesa Research\\ \vspace{0.35cm} \texttt{research@nesa.ai}}}
\date{}

\begin{document}

\maketitle

\section{Introduction}
\label{sec:intro}

Artificial Intelligence (AI), particularly machine learning (ML), has grown significantly in recent decades \cite{jordan2015machine}.
Since the introduction of large language models (LLMs) such as ChatGPT \cite{openai2023chatgpt}, 
Claude \cite{anthropic2023claude}, Gemini \cite{google2023gemini}, and LLaMA \cite{llama2024}, as well as diffusion models \cite{ho2020denoising} like DALLE-3 \cite{openai2023dalle3} and Sora \cite{openai2024sora}, these foundation models have attracted significant interest \cite{bommasani2021opportunities}.
They exhibit advanced capabilities such as in-context learning \cite{brown2020language} and chain-of-thought reasoning \cite{wei2022chain}, yet privacy challenges arise when these models are deployed across distributed or decentralized infrastructures \cite{kairouz2021advances}.

In many scenarios—especially in healthcare \cite{price2019privacy}, finance, or other regulated domains \cite{liu2021privacy}—\textit{data privacy} is a central requirement. Users often need to ensure that sensitive data (e.g., medical images, personal identifiers, or transaction records) are not visible to untrusted nodes in a distributed inference pipeline. 
Existing methods like secure multiparty computation (SMPC) \cite{evans2018pragmatic}, homomorphic encryption (HE) \cite{gentry2009fully}, and differential privacy (DP) \cite{dwork2014algorithmic} can help, but each involves trade-offs in communication overhead, computational latency, or accuracy.

To address these limitations, we propose \textit{Equivariant Encryption} (EE), a technique that enables large-scale model inference on encrypted data while maintaining near-zero performance overhead. 
By transforming internal representations so that the model can operate on ciphertext as if it were plaintext, EE eliminates the high computational costs typically associated with fully homomorphic approaches. In this work, we:

\begin{itemize}
    \item \textbf{Review} background approaches like SMPC, HE, and DP, emphasizing their strengths and shortcomings for large-model inference (\S\ref{sec:dp_background}--\S\ref{sec:he_background}).
    \item \textbf{Introduce} Equivariant Encryption (\S\ref{sec:ee_general}) as a new framework for preserving data confidentiality throughout neural network pipelines.
    \item \textbf{Demonstrate} a decentralized infrastructure example where EE can protect queries and outputs from untrusted nodes (\S\ref{sec:nesa_framework}).
    \item \textbf{Analyze} potential attack vectors and strategies adversaries might employ to invert or compromise EE, and discuss how to counter them (\S\ref{sec:attack_models}).
\end{itemize}

Overall, we show that Equivariant Encryption preserves both the functionality and throughput of large models in distributed or untrusted environments, bridging a gap between security guarantees and practical latency requirements.

\section{Equivariant Encryption: A Middle Ground for Secure Model Inference}

Before detailing our new method, Equivariant Encryption (EE), we will briefly recap three key tools in privacy-preserving data processing—\textit{differential privacy} (DP) (\S\ref{sec:dp_background}), \textit{secure multi-party computation} (SMPC) (\S\ref{sec:smpc_background}), and \textit{homomorphic encryption} (HE) (\S\ref{sec:he_background}). 
DP manages privacy at the dataset level by adding noise, thereby limiting how much an attacker can deduce about any single record, yet does not encrypt intermediate states during inference.
SMPC splits data and computation across multiple participants, reducing exposure but often demanding complex protocols.
HE allows computations on encrypted data at all times, though it can impose substantial overhead and may struggle with non-linear network layers. 
Our Equivariant Encryption (\S\ref{sec:ee_general}) seeks a balanced approach: rather than fully encrypting every component or depending solely on noise or multi-party flows, EE selectively obfuscates crucial internal representations within LLMs and more, retaining strong confidentiality while minimizing performance cost.

\subsection{Background: Differential Privacy (DP)}
\label{sec:dp_background}

Differential Privacy (DP) is a statistical framework designed to protect individual data records in a dataset, while still allowing meaningful aggregate computations or analyses. Formally, let $D$ and $D'$ be two neighboring datasets differing by a single record. A randomized algorithm $\mathcal{M}$ is said to satisfy $(\varepsilon,\delta)$-DP \cite{dwork2006calibrating} if, for any measurable set $S$,
\[
\Pr[\mathcal{M}(D) \in S] \;\le\; e^\varepsilon \,\Pr[\mathcal{M}(D') \in S] + \delta.
\]
Intuitively, altering one individual’s record does not significantly change the distribution of the algorithm’s outputs, thus limiting privacy risks for each participant.

\paragraph{Classical Mechanisms.}
Several mechanisms can ensure DP under different assumptions:
\begin{itemize}
    \item \textbf{Laplace Mechanism:} Injects noise drawn from a Laplace distribution whose scale depends on the function’s sensitivity, thereby hiding individual contributions.
    \item \textbf{Gaussian Mechanism:} Uses Gaussian noise to achieve $(\varepsilon, \delta)$-DP in settings where high-dimensional outputs are required.
    \item \textbf{Exponential Mechanism:} Chooses outputs with probabilities proportional to a utility function, balancing usefulness with DP constraints.
\end{itemize}
Noise level tuning (e.g., the variance of the distribution) controls the trade-off between privacy strength and accuracy.

\paragraph{Practical Considerations and Composition.}
A notable feature of DP is its handling of sequential queries on the same dataset. 
Multiple runs of DP-protected algorithms incur a composed privacy cost, which can be bounded using additive or more refined composition theorems \cite{dwork2014algorithmic}. In machine learning, differentially private stochastic gradient descent (DP-SGD) \cite{abadi2016deep} clips gradients and adds noise at each update, preserving DP at the expense of some accuracy loss—often more pronounced in large-scale models or complex tasks.

\paragraph{Security Model and Limitations.}
DP restricts what can be inferred about any single record by observing the final outputs or aggregated statistics of an algorithm. However, DP does not encrypt intermediate model activations at inference time, leaving room for leaks if raw data are exposed to an untrusted service during predictions.

\paragraph{Connection to EE.}
DP and EE (\S\ref{sec:ee_general}) solve different but compatible facets of privacy. 
While DP reduces the risk of exposing individual training samples through aggregate statistics or model parameters, EE ensures that the inference pipeline itself never processes raw plaintext data. I
n practice, one might train a model with DP for statistical protection of the training set, then deploy EE to keep inference inputs confidential against adversarial observers. This combination can safeguard both training and inference in a layered privacy architecture.

\subsection{Background: Secure Multi-Party Computation (SMPC)}
\label{sec:smpc_background}

Secure Multi-Party Computation (SMPC) is a cryptographic approach that enables multiple parties, each holding private inputs, to compute a joint function without revealing these inputs to one another. 
Formally, suppose there are $n$ parties $\{P_1, P_2, \dots, P_n\}$ with private inputs $x_1, x_2, \dots, x_n$, and they wish to compute a deterministic function
\[
f(x_1, x_2, \dots, x_n) = y,
\]
where $y$ is the output revealed to some or all of the parties, but each $x_i$ remains hidden.

\paragraph{Classical Constructions.}
SMPC can be realized through various protocols, each with different security assumptions and performance characteristics:
\begin{itemize}
    \item \textbf{Yao’s Garbled Circuits:} Originating with Yao \cite{Yao1982}, this approach encrypts a Boolean circuit such that each party learns nothing beyond its own inputs and the final output. 
    \item \textbf{Secret-Sharing Protocols (BGW):} Introduced by Ben-Or, Goldwasser, and Wigderson (BGW) \cite{BGW1988}, each input is split into multiple shares distributed among parties. Intermediate computations proceed on these shares, ensuring no single share reveals the original input.
\end{itemize}
A hallmark of such constructions is that all parties learn the correct final result $y$, while intermediate values remain masked.

\paragraph{Secret Sharing and Arithmetic Operations.}
A common variant of secret sharing is \emph{additive sharing}, where a secret $x$ over a ring $\mathbb{Z}_q$ is divided into $n$ shares $(x_1, x_2, \dots, x_n)$ such that
\[
x = \sum_{i=1}^{n} x_i \quad (\bmod\, q).
\]
Each party receives one $x_i$. Adding two secrets can be done locally on each party’s shares, whereas multiplication often requires additional steps. The BGW model and later protocols such as SPDZ \cite{Damgaard2012} use \emph{multiplication triplets} and integrity checks to allow correct evaluation of products, even in the presence of malicious adversaries.

\paragraph{Security Models.}
SMPC protocols typically consider:
\begin{itemize}
    \item \emph{Semi-Honest Adversaries:} Parties follow the protocol correctly but try to infer extra information from received messages.
    \item \emph{Malicious Adversaries:} Parties can deviate arbitrarily to extract data or alter the outcome.
\end{itemize}
Security proofs guarantee that any subset of corrupted parties learns nothing beyond the legitimate final output.

\paragraph{Practical Considerations.}
While SMPC obviates the need for a fully trusted server, it often introduces higher computational and communication overhead than a single trusted third party \cite{Cramer2015}. Large-scale SMPC can involve frequent message exchanges, especially for complex operations like matrix multiplication in neural networks. Nonetheless, specialized circuit optimizations and precomputation (e.g., random-beaver triplets in SPDZ) have improved the practicality of SMPC for certain machine learning workloads \cite{Mohassel2017}.

\paragraph{Connection to EE.}
Although SMPC conceals inputs from other parties, it does not necessarily hide internal computations from the machine performing those computations. 
By contrast, EE (see \S\ref{sec:ee_general}) encrypts the \emph{internal representations} used within neural network layers. 
In scenarios where partial computations are offloaded to untrusted infrastructure, SMPC ensures data are shared among multiple parties without revealing secrets, and EE obfuscates the intermediate states of the network. Combined, they form a multi-layered approach, with SMPC covering multi-party input privacy and EE preventing visibility into intermediate neural activations or parameters.

\subsection{Background: Homomorphic Encryption (HE)}
\label{sec:he_background}

Homomorphic Encryption (HE) is a cryptographic framework that keeps data encrypted while still allowing meaningful computations on it. 
This capability supports many secure outsourcing and cloud computation scenarios \cite{gentry2009fully,cheon2017homomorphic}, though practical applications often face significant performance challenges. 
Understanding the basics of HE clarifies why EE focuses on a more targeted approach for neural networks.

\paragraph{Motivation and Basic Setup.}
Consider a user with private data $m \in \mathcal{M}$ that must be processed by an untrusted server. Rather than sending $m$ in plaintext, the user encrypts $m$ to produce $c = E(m)$, where 
\[
E: \mathcal{M} \;\rightarrow\; \mathcal{C}.
\]
The server then operates on $c$ to yield some output $\tilde{c}$. Crucially, the homomorphic property ensures:
\[
D\bigl(f'(c_1, c_2, \dots)\bigr) 
= 
f\bigl(D(c_1), D(c_2), \dots\bigr),
\]
where $D$ is the corresponding decryption function, $f(\cdot)$ is the desired plaintext operation, and $f'(\cdot)$ is its encrypted analog. This principle lets the server process encrypted data without learning $m$ \cite{rivest1978method,paillier1999public}.

\paragraph{Types of HE Schemes.}
HE systems are commonly categorized by how many operations on ciphertexts they support:
\begin{itemize}
    \item \textbf{Partial HE (PHE):} Permits repeated use of one operation—addition or multiplication. For instance, RSA-based schemes support multiplicative homomorphism \cite{rivest1978method}, whereas the Paillier cryptosystem supports additive homomorphism \cite{paillier1999public}.
    \item \textbf{Somewhat or Leveled HE:} Allows both addition and multiplication up to a certain depth, controlled by noise management. This depth determines how many multiplied ciphertexts can be handled before decryption becomes invalid.
    \item \textbf{Fully HE (FHE):} Provides unlimited additions and multiplications, often through “bootstrapping” to periodically refresh ciphertexts and limit noise \cite{gentry2009fully}.
\end{itemize}

\paragraph{Ring-Based Construction and Polynomial Representation.}
Modern FHE schemes (e.g., BFV \cite{fan2012somewhat}, CKKS \cite{cheon2017homomorphic}) typically use polynomial rings for computational efficiency. A cyclotomic polynomial ring
\[
\mathcal{R} = \mathbb{Z}_q[x]\big/\langle x^N + 1 \rangle
\]
serves as the plaintext space, with additional polynomials denoting ciphertexts. Security derives from adding controlled ``noise'' that grows with each operation. If not managed, excessive noise can invalidate decryption.

\paragraph{Computational Overheads and Trade-Offs.}
Despite extensive research and optimizations, HE can remain much more resource-intensive than plaintext processing \cite{albrecht2019homomorphic}. 
Ciphertext sizes and polynomial arithmetic introduce overhead, and advanced batching or leveled HE schemes \cite{halevi2014alg} partially mitigate but do not eliminate these costs. 
In particular, LLMs or deep neural architectures demand numerous matrix multiplications across many layers, challenging HE’s performance in real-time or large-scale settings. Parameter tuning, relinearization, and ciphertext expansion can increase both latency and memory usage.

\paragraph{Connection to EE.}
EE leverages the concept of secure computation over transformed data but confines encryption to certain high-risk network layers, rather than fully encrypting the entire computational graph. 
By restricting complex or noise-sensitive operations to plaintext, EE dramatically reduces the overhead commonly associated with HE, yet still prevents exposure of critical internal representations. 
As we discuss in the next sections (\S\ref{sec:ee_general}), this selective encryption yields a more manageable trade-off between runtime performance and data confidentiality in modern neural networks.

\subsection{Equivariant Encryption: A Practical Solution for Blind Inference}
\label{sec:ee_general}

Equivariant Encryption (EE) is presented here as a selective encryption technique for neural network inference, avoiding the high overhead of fully HE and circumventing the limitations of trusted execution environments (TEEs) or DP. 
EE keeps inputs and outputs confidential while preserving near-zero additional latency, making it suitable for large-scale models or time-critical applications.

\begin{figure}[!ht]
  \centering
  \includegraphics[width=0.75\textwidth]{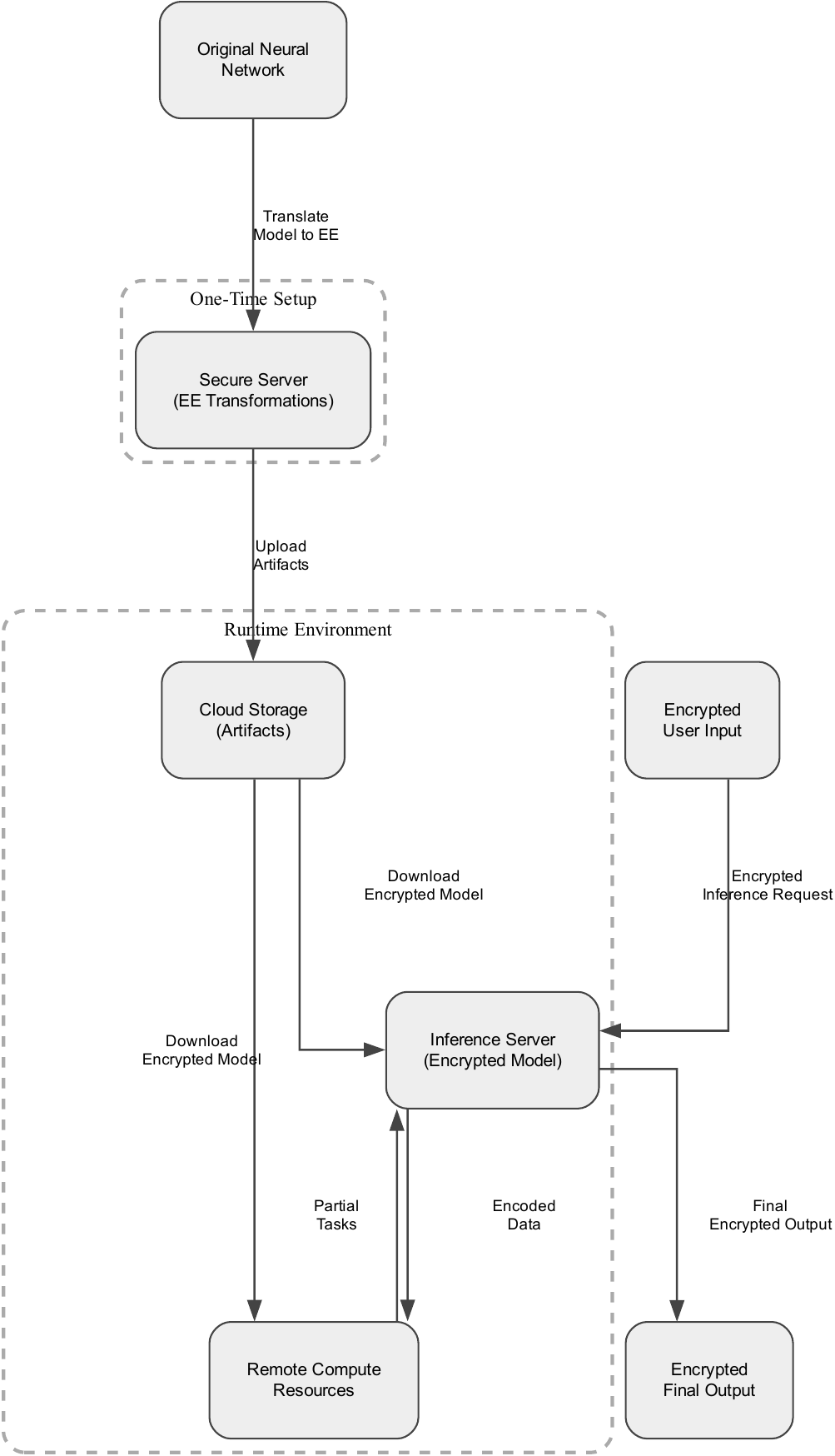}
  \caption{%
    A concise illustration of Equivariant Encryption’s workflow. 
    A one-time setup (\emph{top}) applies EE transformations to the model on a secure server,
    and the runtime environment (\emph{middle}) handles encrypted model artifacts
    along with user queries. This ensures requests and responses remain unreadable 
    by any untrusted infrastructure.
  }
  \label{fig:ee_diagram}
\end{figure}

\paragraph{Key Characteristics and Advantages.} EE has the following advantages:
\begin{itemize}
    \item \textbf{Complete Server Blindness:} In an EE-based pipeline, raw data, queries, and intermediate activations never appear in plaintext on the server.
    \item \textbf{Negligible Latency:} EE sidesteps the typical performance pitfalls of full HE, allowing inference speeds comparable to standard unencrypted processing.
    \item \textbf{Broad Model Applicability:} From CNNs to LLMs with attention blocks, EE can accommodate a variety of deep-learning architectures, including multi-modal pipelines.
    \item \textbf{Cost-Effectiveness:} By eliminating the need for specialized hardware (as in TEEs) or complex parameter setups (as in HE), EE can lower operating expenses for on-prem or cloud-based deployments.
    \item \textbf{RAG and Beyond:} Retrieval-augmented generation workflows remain encrypted end to end, preserving both queries and retrieved documents from external inspection.
    \item \textbf{Simple Integration:} EE typically requires minimal changes in code, such as replacing specific layer operations with “encrypted” equivalents. 
\end{itemize}

\paragraph{Motivation: ``Blind AI'' Without Performance Penalties.}
Safeguarding privacy during inference poses significant challenges, particularly for large-scale models and real-time systems. Existing methods have notable drawbacks:
\begin{itemize}
    \item \textit{HE:} Encrypts all operations but struggles with non-linear layers and can incur large runtime expansions.
    \item \textit{TEEs:} Rely on hardware trust, granting potential backdoor privileges to system administrators.
    \item \textit{DP:} Obscures individual contributions through noise but may not secure intermediate activations from a malicious inference server.
\end{itemize}
Equivariant Encryption addresses these gaps by focusing on layer-specific transformations, retaining strong data confidentiality with minimal overhead.

\paragraph{Overview of EE.}
EE works by converting data and selecting neural operators into a specialized “encrypted domain” (Figure~\ref{fig:ee_diagram}).
Rather than encrypting every operation via polynomial-based homomorphisms, EE tailors transformations to each layer’s structure. This customization permits the network to handle encrypted vectors nearly as if they were plaintext, without the computational blowup seen in fully homomorphic approaches.

Formally, we have the following definition for EE:
\begin{definition}[Equivariant Encryption]
Given any plaintext $p$, EE is an encrypt-decrypt algorithm such that
\begin{itemize}
\item
Recoverability:
\begin{equation}
p=\mathrm{decrypt}(\mathrm{encrypt}(p)),
\end{equation}
\item
Equivariance:
\begin{equation}
\mathrm{decrypt}(F(\mathrm{encrypt}(p))) = F(p),
\end{equation}
\end{itemize}
where $F$ represents any linear operations and a specific set of supported non-linear operations.
\end{definition}
Currently, our framework directly supports the following set of activation and processing functions: ReLU, GeLU, SiLU, RMS Normalization, and Layer Normalization. The framework can also support other non-linear functions without requiring any modifications.


\paragraph{Comparison with HE.}
While both EE and HE enable computations on encrypted data, they differ in overhead and flexibility:

\begin{table}[h]
\centering
\caption{Equivariant Encryption (EE) vs.\ Homomorphic Encryption (HE) for Neural Inference.}
\label{tab:ee_vs_he}
\scalebox{0.84}{
\begin{tabular}{lcc}
\toprule
\textbf{Property} & \textbf{EE} & \textbf{HE} \\
\midrule
\textbf{Latency Overhead} & Near-zero & High \\
\textbf{Handling of Non-linear Ops} & Exact & Often approximations \\
\textbf{Key Management} & User-defined & Tied to HE scheme \\
\textbf{Security Basis} & Large combinatorial space & Lattice / number theory \\
\textbf{Scalability to Large Models} & Straightforward & Resource-intensive \\
\textbf{Accuracy} & Matches plaintext & Potential approximation loss \\
\textbf{Integration Complexity} & Layer-by-layer transforms & Major re-engineering \\
\bottomrule
\end{tabular}}
\end{table}

\paragraph{EE in Practice: Minimal Overheads and Realistic Security.}
All transformations are applied once, offline, ensuring the final “encrypted model” maintains the same order of multiplications and additions as an unencrypted version. Consequently, runtime latencies mirror those of plaintext inference. Compromising the data would require inverting $T$—frequently a high-dimensional transform—rendering brute-force or direct linear-algebraic attacks computationally infeasible.

\paragraph{Deployment Scenarios.}
\begin{itemize}
    \item \textit{LLMs and Conversational Systems:} Token embeddings become encrypted embeddings so no plaintext tokens ever appear on the server.
    \item \textit{Vision Models:} Encrypted feature maps flow through convolution and activation layers with minimal overhead.
    \item \textit{RAG Pipelines:} Queries and retrieved content remain enciphered, preventing servers from inspecting user context or knowledge sources.
\end{itemize}

\paragraph{Summary.}
Equivariant Encryption represents a pragmatic, high-performance alternative to fully homomorphic encryption for blind inference. By using a selective approach, encrypting only layers at the highest risk of leaking information, EE achieves robust privacy without sacrificing speed. In large-scale deployments, from LLM serving to real-time analytics, it provides a compelling solution for “always-encrypted” inference that remains both practical and secure.

\subsection{Use Case: A Decentralized Infrastructure Example of EE}
\label{sec:nesa_framework}

\begin{figure}[!t]
  \centering
  \includegraphics[width=1.0\columnwidth]{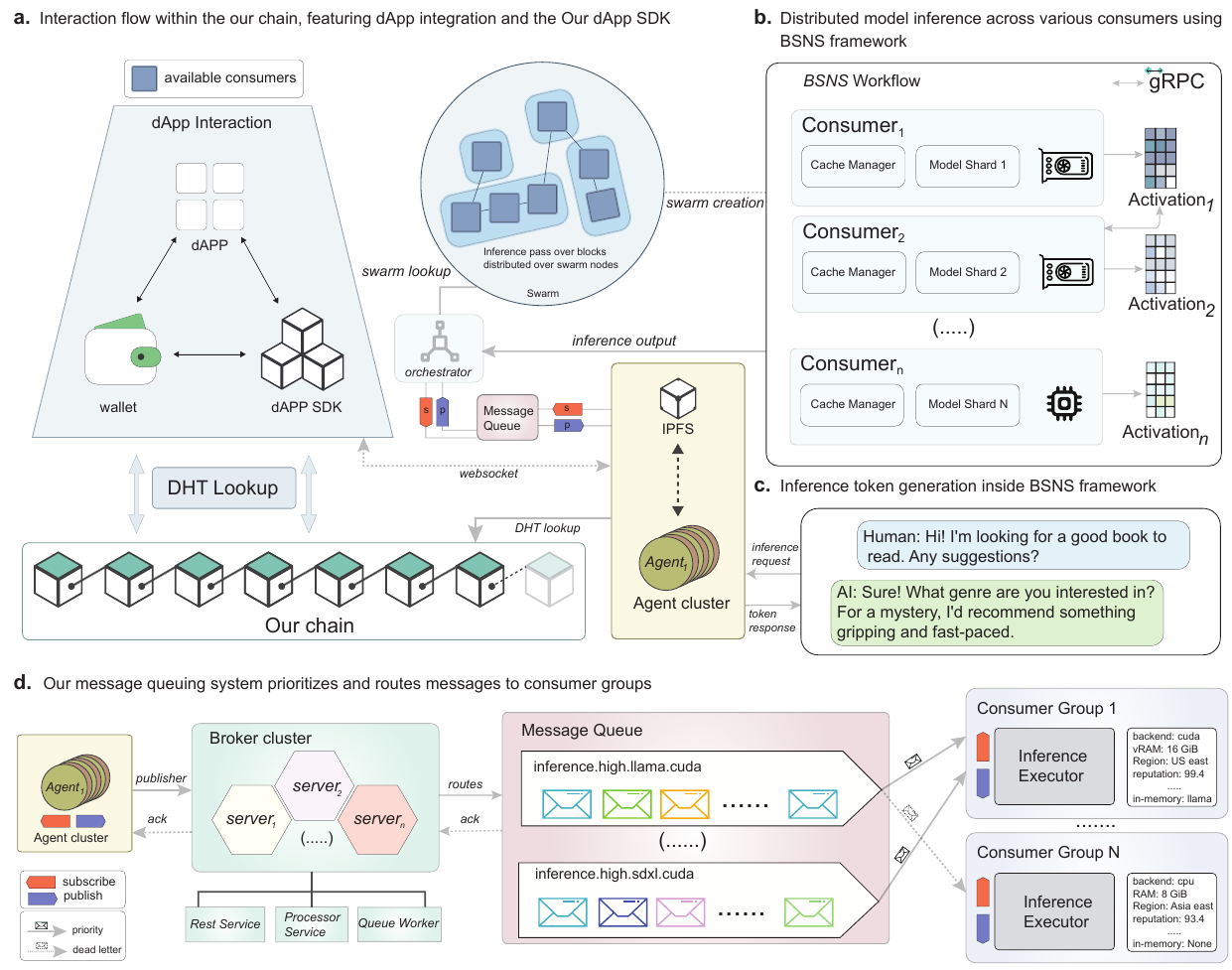}
  \caption{\textbf{System Overview.} 
  This figure shows one example of a decentralized inference flow for large-scale models.
  \textbf{(a)} Depicts the main interaction path: from a dApp and wallet to a distributed hash table (DHT) lookup across the chain. 
  \textbf{(b)} Highlights a framework that splits a large model into shards for parallel processing, passing activations via gRPC. 
  \textbf{(c)} Demonstrates a text-generation query: an agent checks chain transactions and returns the result to the user-facing dApp. 
  \textbf{(d)} Shows a message queuing system that assigns requests to different consumer groups based on resource constraints, reputation scores, and model demands.}
  \label{fig:bsns_architecture}
\end{figure}

Although EE applies broadly to any scenario requiring private model inference, this section presents a concrete decentralized infrastructure example, inspired by frameworks that split model execution among multiple nodes or shards. 
Figure~\ref{fig:bsns_architecture} illustrates a setting where:
\begin{itemize}
    \item A query enters the system through a decentralized application (dApp) \cite{zheng2020overview} and a wallet mechanism.
    \item The query, along with relevant state, is dispatched across a blockchain-like infrastructure, performing a distributed hash table (DHT) lookup for transactions.
    \item A message-broker subsystem manages job routing to multiple nodes, each responsible for processing a portion (shard) of a large model \cite{rajbhandari2020zero}.
    \item Activations and partial outputs flow through gRPC-based links, and final results are stitched together for the user.
\end{itemize}

Such distributed systems are attractive for scalability and fault-tolerance but can raise privacy questions: 
intermediate activations, user queries, or model outputs may be visible to untrusted parties at each node. 
\emph{Equivariant Encryption} addresses this challenge by encrypting the internal representations, ensuring that no node—except the original client—can interpret the raw data or glean sensitive information. 
As described in \S\ref{sec:ee_general}, EE focuses on carefully chosen transformations that maintain the correctness of computations while preventing adversaries from reconstructing user inputs or outputs. 
In this sense, it complements existing decentralized methods by preserving high performance without sacrificing privacy.

\section{Threat Analysis and Attack Models}
\label{sec:attack_models}

Having introduced EE as a general technique for secure model inference, we now focus on potential attacks against such systems. 
This section formalizes how attackers might attempt to invert or bypass EE when data are transmitted (and processed) in an encrypted form. Although the following examples refer to a network context inspired by decentralized inference and token-based LLM protocols, these considerations apply broadly wherever EE is used to conceal intermediate representations or token mappings.

\subsection{Attack Vector Background}
\label{subsec:attack_vector}

We focus on the scenario in which requests and responses are transmitted via HTTP in an \emph{equivariantly} encrypted form. Specifically, the tokens that represent inputs and outputs for a large language model (LLM) are permuted or transformed according to an unknown mapping. Bad actors intercepting these encrypted token IDs gain access only to a transformed sequence; the legitimate user or trusted client alone knows the key(s) or mapping required to recover the original token IDs. 

For concreteness, assume the attacker obtains input-output pairs over some duration. Each pair $(I_i, O_i)$ is represented by sequences of token IDs that have been scrambled through EE. The adversary’s goal is to reconstruct or guess the original plaintext tokens used by the standard tokenizer. This setting highlights the difference between \emph{observing} encrypted token sequences and actually \emph{inverting} them. 

\subsection{A Unified Analytical Framework}
\label{subsec:attack_framework}

To systematically study potential attacks, we consider a mathematical optimization viewpoint. Consider a target LLM, such as a Llama-family model, which implements a function 
\[
f: \mathcal{C} \to \mathcal{C},
\] 
mapping a token-sequence input in some dictionary $\mathcal{V}$ (where $|\mathcal{V}|$ may be up to 128K tokens) to a token-sequence output. Depending on the sampling mechanism, $f$ can be deterministic (greedy decoding) or stochastic (temperature-based or top-$k$ sampling).

An attacker observes $n$ pairs of encrypted input-output sequences $\{(I_i,O_i)\}_{i=1}^n$, with each $I_i, O_i \in \mathcal{C}$ after scrambling by EE. The adversary knows the vocabulary set $\mathcal{V}$ but not the specific permutation or mapping $\mathbb{P}$ that recovers plaintext tokens. To mount an attack, the adversary tries to find a mapping $\mathbb{P}: \mathcal{V} \to \mathcal{V}$ such that:
\[
\mathbb{P}(O_i) \;\approx\; f\bigl(\mathbb{P}(I_i)\bigr), 
\]
and the decrypted sequences $\bigl(\mathbb{P}(I_i), \mathbb{P}(O_i)\bigr)$ form a semantically valid question-answer or prompt-response pair. Formally, one might frame this as:
\begin{equation}
\label{equ:framework}
    \min_{\mathbb{P}} \;\;\frac{1}{n}\sum_{i=1}^n \mathcal{L}\bigl(\mathbb{P}(I_i),\mathbb{P}(O_i)\bigr)
    \quad 
    \text{s.t.}
    \quad 
    \mathbb{P}(O_i)=f\bigl(\mathbb{P}(I_i)\bigr)\; \forall i,
\end{equation}
where $\mathcal{L}\bigl(\cdot,\cdot\bigr)$ is a loss function that captures how well the decrypted pairs match valid natural language usage and plausible model responses.

\paragraph{Challenges.} We witness the following challenges for solving Equation \eqref{equ:framework}:
\begin{itemize}
    \item \textbf{Loss Function Design:} What semantic or linguistic constraints best reflect the adversary’s prior knowledge? For instance, knowledge of frequency distribution (e.g., tokens like “the,” “of,” “and” occur frequently) or grammar structure might be integrated into $\mathcal{L}$.
    \item \textbf{Discrete Optimization:} Finding a permutation $\mathbb{P}$ that satisfies the above constraints is a high-dimensional combinatorial problem on the order of $|\mathcal{V}|!$, which is intractable to solve exactly for large vocabularies.
\end{itemize}

\subsection{Baseline Attacks}
\label{subsec:baseline_attacks}

In practice, adversaries often resort to heuristic or partial methods for solving \eqref{equ:framework}. Below, we outline several baseline approaches.

\subsubsection{Designing a Loss Function}
\label{subsubsec:loss_design}

\paragraph{LLM-as-a-Judge.}
One concept is to leverage a powerful reference model (e.g., GPT-4 or another advanced LLM) to score how consistent a decrypted output $\mathbb{P}(O_i)$ is with the corresponding input $\mathbb{P}(I_i)$. For instance, the attacker can prompt the reference LLM to rate the coherence or correctness of the response from 0 to 10, assigning a lower loss for better Q\&A alignment. This approach effectively uses a large model’s own understanding to guess whether a proposed permutation is valid.

\paragraph{Linguistic Domain Knowledge.}
Alternatively, the adversary can incorporate domain expertise or statistical cues. For example, the frequency of certain tokens (e.g., “the,” “is,” “and”) might be recognized in plaintext language, and grammar rules (e.g., subject-verb-object sequences) can guide guesses about which tokens appear in typical positions. These heuristics inform $\mathcal{L}\bigl(\mathbb{P}(I_i),\mathbb{P}(O_i)\bigr)$ to penalize permutations that fail to produce plausible word frequencies or syntactic structures.

\subsubsection{Designing an Optimizer}
\label{subsubsec:optimizer_design}

Even with a well-defined $\mathcal{L}$, solving for a global or local minimum in \eqref{equ:framework} can be difficult. We outline three heuristic attacks:

\paragraph{Brute Force.}
The naive method enumerates \textit{all} permutations of $\mathcal{V}$, computing the loss each time. With complexity $|\mathcal{V}|!$, this is clearly infeasible beyond very small vocabularies.

\paragraph{Random Sampling.}
A more tractable (though still limited) approach randomly draws $M$ permutations from the space. The attacker then evaluates $\mathcal{L}$ and chooses the lowest-loss candidate. Genetic algorithms or other population-based methods can improve upon pure random sampling by “breeding” permutations that yield better fitness scores.

\paragraph{Hill-Climbing.}
Starting from a random or heuristic permutation, an attacker iteratively searches for local improvements by swapping two token mappings at a time. If a swap lowers $\mathcal{L}$, the permutation is updated. This process continues until no improving swaps are found or computational resources are exhausted. While the algorithm may get stuck in local minima, it can be more effective than random guessing for moderate vocabulary sizes.

\paragraph{Summary.}
These baseline attacks demonstrate how an adversary might attempt to invert or weaken Equivariant Encryption by exploiting partial linguistic cues or iterative search heuristics. In large-scale LLM scenarios—with extensive vocabularies and highly varied text inputs—the complexity of inverting the token transformations remains considerable. Nonetheless, these methods highlight the importance of carefully choosing transformations and ensuring sufficient dimensional and combinatorial complexity in EE, so that feasible attacks remain prohibitively expensive in practice.

\section{Benchmarking}

\subsection{Language Models}

\subsubsection{Fidelity Score}
The \textit{fidelity score} measures the similarity of confidence values for the generated logits between two inference runs. It is defined as:

\begin{equation}
    \text{Fidelity} = 1 - \frac{\sum_{i=1}^{n} \frac{|s_i^{EE} - s_i^{VI}|}{\max(s_i^{EE}, s_i^{VI})}}{n},
\end{equation}

where:
\begin{itemize}
    \item \( n \) is the total number of samples.
    \item \( s_i^{VI} \) and \( s_i^{EE} \) are the class/first token confidence scores for the \( i \)-th sample from Vanilla Inference (VI) and Equivariant Encryption (EE), respectively.
\end{itemize}

A higher fidelity score indicates that the EE model produces confidence values that are more similar to the VI model.  Our benchmarking for text models is as follows. For IMDB dataset, we sampled 5000 entries. For LLMs, we used MT-Bench plus 2000 entries sampled from ShareGPT repeated twice. 


\begin{table}[h]
    \centering
    \renewcommand{\arraystretch}{1.3}
    \caption{Comparison of inference latency, fidelity, and output consistency between Vanilla Inference (VI) and Equivariant Encryption (EE) across various language models. Latency is measured in seconds, while fidelity is evaluated on IMDb (for BERT models) and ShareGPT (for LLMs). Standard deviation in inference time is also provided as a percentage in inference time.}
    \scalebox{0.74}{
    \begin{tabular}{lll|cccccc}
        \toprule
        \textbf{Model} & \textbf{vLLM?} & \textbf{bs} & \textbf{VI (s)} & \textbf{EE (s)} & $\Delta$\textbf{T (\%)} & \textbf{Fid (\%)} & $\Delta$\textbf{T Std (\%)}  \\
        \midrule
        BERT-base & No & 1 & 39.02 & 39.27 & \(-0.64\%\) & 92.38 & \(\pm 0.89\%\) \\
        Sentiment-BERT & No &  1 &  39.16 & 39.08 & \(+0.20\%\) & 88.35 & \(\pm 0.91\%\) \\
        RoBERTa-base & No & 1 & 40.31 & 39.98 & \(+0.82\%\) & 99.85 & \(\pm 0.87\%\) \\
        Llama 3.1-8B & No & 1 & 418.58 & 455.68 & \(+8.88\%\) & 99.999 & \(\pm 0.78\%\)\\
        Llama 3.1-8B & Yes & 256 &  293.88 & 292.33 & \(-0.53\%\) & 99.999 & \(\pm 1.18\%\) \\
        \bottomrule
    \end{tabular}
    }
    \label{tab:inference_comparison}
\end{table}

\bibliographystyle{unsrt}
\bibliography{references}

\end{document}

%% file: dfn.tex
\definecolor{cobalt}{rgb}{0.03, 0.27, 0.49}
\definecolor{azurecolorwheel}{rgb}{0.0, 0.5, 1.0}
\definecolor{cadmiumgreen}{rgb}{0.0, 0.42, 0.24}